\documentstyle[12pt]{article}
  \def\thebibliography#1{{\bf{References}}\list
 {[\arabic{enumi}]}{\settowidth\labelwidth{[#1]}\leftmargin\labelwidth
   \advance\leftmargin\labelsep
   \usecounter{enumi}}
   \def\newblock{\hskip .11em plus .33em minus -.07em}
   \sloppy
   \sfcode`\.=1000\relax}
  
\begin{document}
\def\ttb{t\bar t}
\def\qqb{Q\bar Q}
\def\gg{\gamma\gamma}
\def\bt{\beta_{t}}
\def\to{\rightarrow}
\def\RM{{R(0)^2\over 2\pi M}}
\def\MeV{{\rm MeV}}
\def\Lms#1{\Lambda_{\overline{MS}}^{(#1)}}
\def\asp{{\alpha_s\over\pi}}
\def\ams{\alpha_{\overline {MS}} }
\def\amsp{{\ams \over\pi}}
\def\eps{\epsilon}
\newcommand{\hoch}[1]{\mbox{\rule[0cm]{0cm}{#1}}}
\everymath={\displaystyle}
\thispagestyle{empty}
\vspace*{-2mm}
\thispagestyle{empty}
\noindent
\hfill TTP93--10\\
\mbox{}
\hfill  March  1993  \\   
\vspace{0.5cm}
\begin{center}
  \begin{Large}
  \begin{bf}
EXOTIC BOUND STATE PRODUCTION \\
AT  HADRON COLLIDERS\footnote{\normalsize Supported by
 BMFT Contract 055KA94P}
   \\
  \end{bf}
  \end{Large}
  \vspace{0.8cm}
  \begin{large}
   J.H. K\"uhn and E.\ Mirkes\\[5mm]
    Institut f\"ur Theoretische Teilchenphysik\\
    Universit\"at Karlsruhe\\
    Kaiserstr. 12,    Postfach 6980\\[2mm]
    7500 Karlsruhe 1, Germany\\
  \end{large}
  \vspace{4.5cm}
  {\bf Abstract}
\end{center}
\begin{quotation}
\noindent
Hadronic production of nonrelativistic boundstates of $b^{\prime}$ or
isosinglet quarks D with suppressed weak decays is investigated
for LHC and SSC energies.
QCD corrections to production and decay rates are incorporated.
Large rates for final states from $\eta_{b^{\prime}}\rightarrow HZ$
are predicted.
\end{quotation}
\newpage
\thispagestyle{empty}
\mbox{}
\newpage
\setcounter{page}{1}
The theoretical and experimental aspects of hadronic
quarkonium production have attracted a great deal of
interest after the original discovery of $J/ \Psi$ and
$\Upsilon$.  Gluon
fusion seemingly dominates the cross section with a small
admixture of quark-antiquark annihilation in $p\bar{p}$ and
$\pi^{-}p$
collisions at lower energies.  Given sufficiently high
center of mass energy and luminosity, the same parton
processes could also lead to toponium production at a future
hadron collider.  The production rate of $1^{--}$ bound states
through the reaction
$g+g\rightarrow 1^{--}+g$
 is tiny.  The cross section for
gluon-gluon fusion into the ${}^{1}S_{0}$  state $\eta_{t}$,
 however, would be
sufficiently large to produce a large number of resonances.
The experimental signal
that might conceivably be accessible
under realistic assumptions
is the $\gamma\gamma$ decay
mode of $\eta_{t}$ \cite{Rubbia}. However, this mode
 is severely reduced by the tiny branching ratio of
order $10^{-4}$ down to $10^{-5}$
--- a consequence of the large single
quark decay rate of $\eta_{t}$.
Detailed investigations which include
higher order corrections for the production rate and a
realistic QCD potential show that the window for the
discovery of $\eta_{t}$ is extremely narrow.
Fairly optimistic
assumptions have to be employed to access quark masses even
in the range of 100 to 110 GeV \cite{letter,rev}.

However, the situation changes drastically if the single
quark decay mode
of the ${}^{1}S_{0}$ bound state
is absent or at least strongly suppressed,
 as is the case for a $b^{\prime}$ with
suppressed couplings to the lighter u, c, and t quarks, or
 for
isosinglet quarks denoted in the following by $D$.
 The branching ratio for decay modes of
these bound states with truly characteristic
final states such as
$\gamma\gamma,\, \gamma Z,\,ZZ,\, WW,\,HZ$
are no longer drastically suppressed and may
even become dominant, as is the case
for the $HZ$ mode in the high mass region.
  These would be easily
accessible in future experiments at LHC or SSC, for quark
masses even up to 1000 GeV.  Predictions for these reactions
in Born approximation have been derived in references
\cite{acta,barger,kurep}.

In a recent paper \cite{letter,rev},
QCD corrections were evaluated for the
production of nonrelativistic ${}^{1}S_{0}$
bound states in hadronic
collisions.  These corrections
 are positive and fairly
large, justifying the renewed investigation of exotic bound
state production.
The corrections
 can be also applied
to the production of bound states
composed of a $b^{\prime}$ with $I_{3}=-{1}/{2}$
and $e_{Q}=-{1}/{3}$ or of a $D$ quark with $I_{3}=0$ and
$e_{Q}=-{1}/{3}$.

In the following we specify
the general structure of the cross section for
$\eta_{b',D}$ production in hadronic collisions
\begin{equation}
h_{1}(P_{1})+h_{2}(P_{2})\rightarrow\eta_{b',D}+X
\label{hadwq}
\end{equation}
in the framework of perturbative QCD.
In (\ref{hadwq})
 $h_{1}$ and $h_{2}$ denotes unpolarized hadrons with
momenta $P_{i}$.
The hadronic cross section in NLO is thus given by
\begin{equation}
\sigma^{H}(S)
=\int dx_{1} dx_{2}
f_{a}^{h_{1}}(x_{1},Q^{2}_{F})f_{b}^{h2}(x_{2},Q^{2}_{F})
 \hat{\sigma}^{ab}(s=x_{1}x_{2}S,
 \alpha_{s}(\mu^{2}),\mu^{2},Q^{2}_{F})
\end{equation}
where one sums over $a,b=q,\bar{q},g$.
$f_{a}^{h}(x,Q^{2}_{F})$
is the probability density to find parton $a$ with
fraction $x$
in hadron $h$ if it is probed at scale $Q_{F}^{2}$
and $\hat{\sigma}^{ab}$
 denotes the parton cross section for the process
\begin{equation}
a(x_{1}P_{1})+b(x_{2}P_{2})\rightarrow \eta_{b',D}+X
\end{equation}
from which collinear
initial state singularities have been factorized out at a
scale $Q_{F}^{2}$
and implicitly included in the scale-dependent parton
densities $f_{a}^{h}(x,Q_{F}^{2})$.
The partonic cross sections  $\hat{\sigma}^{ab}$ in NLO
 for the $gg, qg$ and $q\bar{q}$ initiated
 reactions are given in \cite{rev}.  They
are proportional to the square of the bound state
 wave function at the
origin $R(0)$
and hence dependent on the QCD potential.  For the
evaluation we employ the two-loop QCD potential $V_{J}$
\cite{Igi}.
 The
numerical results depend critically on the value of the QCD
parameter $\Lms{4}$.
The results for the dimensionless quantity
 $|R(0)|^{2}/M^{3}_{\eta}$ ($M_{\eta}=2m_{b^{\prime,D}}$) are
displayed in
 fig. 1
 for different values of $\Lms{4}$
(200, 300 and 500 MeV)
 to indicate the characteristic
uncertainty in the prediction.
These $\Lambda$ values
correspond to
$\alpha_{s\,\overline{MS}}^{(5)}
=0.109,\, 0.1165$ and 0.127 respectively
 (and to
 $\Lms{5}= 0.132,\, 0.207$ and 0.366 MeV).
  For the parton distribution functions
we employ MT set B1 with $\Lms{4}=194\, \mbox{MeV}$ \cite{MT} and
work in the DIS factorization
scheme.  The
renormalization and factorization scales are fixed
at $M=M_{\eta}$.
We use the $\overline{{MS}}$ definition
 of $\alpha_{s}$ at two-loop
accuracy with six flavours
\begin{equation}
\alpha_{s\,\overline{MS}}(Q)=
\frac{12\pi}{(33-2n_{f})\ln(Q^{2}/\Lambda^{2})}
\left[1-\frac{6(153-19n_{f})}{(33-2n_{f})^{2}}
\frac{\ln\ln(Q^{2}/\Lambda^{2})}{\ln(Q^{2}/\Lambda^{2})}\right]
\end{equation}
 and
 $\Lms{6}=58$ MeV to be consistent with the
$\Lms{4}$ value from the parton distributions.

  The $K-$factor and the
production cross section
 are shown in
figs. 2 and 3a respectively
 as functions of the bound state mass for three
different $\Lambda$  values in the potential.
  The
uncertainty from the choice of the parton distribution
functions is shown in fig. 3b.
  The additional
uncertainty from the choice of
the renormalization and factorization scales is estimated to
be less than 10\% \cite{rev}.

To obtain predictions for the interesting reactions
these production cross sections have to be multiplied by the
various possible branching ratios.
It is convenient to normalize the decay rates relative
to the two photon decay rate $\Gamma_{\gamma\gamma}$
with
\begin{equation}
\Gamma_{\gamma\gamma}=12e_{Q}^{4}\alpha^{2}\frac{|R(0)|^{2}}{M^{2}}
\label{gamborn}
\end{equation}
With $R_{ab}:=\Gamma_{ab}/ \Gamma_{\gamma\gamma}$
($ab=\gamma Z,\,HZ,\,ZZ,\,WW,\,gg,\,t\bar{t})$
one finds
\newpage
\begin{eqnarray}
R_{\gamma Z}&=&
2\,\frac{v_{Q}^{2}}{e_{Q}^{2}y^{2}}
\left(\frac{2p}{M}\right)
\label{dgamz}
\\[2mm]
R_{HZ} &=&
\,\,\,\frac{a_{Q}^{2}}{e_{Q}^{4}}\frac{1}{\rho_{Z}^{2}y^{4}}
\left(\frac{2p}{M}\right)^{3}
\label{dhz}
\\[2mm]
R_{ZZ} &=&
\,\,\,\frac{(v_{Q}^{2}+a_{Q}^{2})^{2}}{e_{Q}^{4}y^{4}}
\frac{1}{(1-2\rho_{Z})^{2}}
\left(\frac{2p}{M}\right)^{3}
\label{dzz}
\\[2mm]
R_{WW} &=&
\,\,\,\frac{1}{2e_{Q}^{4}(4\sin^{2}\theta_{W})^{2}}
\frac{1}{(1-\rho)^{2}}
\left(\frac{2p}{M}\right)^{3}
\label{dww}
\\[2mm]
R_{gg}&=&
\frac{2}{9}\,\frac{\alpha_{s}^{2}(M)}{
e_{Q}^{4}\alpha^{2}}
\label{dgg}
\\[2mm]
R_{t\bar{t}}&=&
6\,\frac{a_{Q}^{2} a_{t}^{2}}{e_{Q}^{4}y^{4}}
\frac{m_{t}^{2}M^{2}}{M_{Z}^{4}}
\left(\frac{2p}{M}\right)
\label{dtt}
    \end{eqnarray}
where
\begin{equation}
  \begin{array}{ll}
\rho_{i}\equiv M_{i}^{2}/M^{2}\,\, ,\hspace{1cm} &
\rho\equiv 2\left(1/4-\rho_{Q^{\prime}}+\rho_{W}\right)\,\, ,\\[2mm]
v_{Q}\equiv 2I_{3\,Q}-4e_{Q}\sin^{2}\theta_{W}\,\,,&
a_{Q}=2 I_{3\,Q} \hspace{1cm}
y\equiv 2\sin 2\theta_{W}
\label{abb}
  \end{array}
\end{equation}
QCD corrections to the decay rates  will be discussed below.
In eq. (\ref{abb}) $I_{3\,Q}$ denotes the third component
 of the weak isospin
of
the $Q-$type quark and $\theta_{W}$ is the Weinberg angle.
$p$ denotes the three-momentum of the final state particles.
The single quark decay mode which dominates for $\eta_{t}$ proceeds
through the decay of the heavy quark into  $b+W$.  For a heavy
sequential quark $b^{\prime}$ of isospin $I_{3}=-{1}/{2}$ the analog
decay $b^{\prime}\rightarrow t^{\prime}+W$ is presumably strongly
suppressed by a small
mixing angle $\theta^{\prime}$
 if the mass of the isospin partner $t^{\prime}$ is
larger than $m_{b^{\prime}}$.
Once ${\theta^{\prime}}^{2} < 10^{-3}$
 the annihilation decays dominate
and single quark decays can be ignored for most practical purposes.
 Similar
considerations apply to decays of isosinglet quarks $D$ into
$b+Z, t+W$ or $c+W$
which are also inhibited by small mixing angles.  For
the subsequent discussion these single quark modes will be
ignored.  The dominant decays are thus all
 proportional to the
wave function squared, the branching ratios
$\Gamma_{ab}/\sum\Gamma_{ab}$
 are independent
of the wave function, and as a result can be predicted
unambiguously.

The WW mode needs further specification:
The decay of $\eta_{b^{\prime}}$  into $WW$
 proceeds through the virtual isospin
partner $t^{\prime}$ and hence depends on the mass of $t^{\prime}$.
Present limits on the deviation of the $\rho-$parameter from one
as deduced from electroweak precision measurements limit the
mass splitting $\bigtriangleup m $
 between $m_{t^{\prime}}$ and $m_{b^{\prime}}$ in a drastic way and
 $\bigtriangleup m < 200 \,\,\mbox{GeV}$
can be considered a generous upper bound.
For simplicity we shall assume equal $b^{\prime}$ and $t^{\prime}$
masses
($\rho_{Q^{\prime}}=1/4$ in eqs. (\ref{dww},\ref{abb}))
in the subsequent
numerical evaluation.

For bound states $\eta_{D}$ from isosinglet
quarks  the WW decay may in principle
proceed through virtual quarks of charge 2/3.  However,
the $D$-$t$-$W$-coupling is suppressed by small mixing angles and
the WW mode can therefore safely be neglected.
The $HZ$ mode is absent as a consequence of $a_{D}= 0.$

QCD corrections to annihilation decays
are at present available  for hadronic  ($gg-$),
for  $\gamma\gamma-$  and for the $t\bar{t}-$decay modes.
\begin{eqnarray}
\Gamma_{\gamma\gamma}&=&
\Gamma_{\gamma\gamma}^{Born}
\left[1+\frac{\alpha_{s}(\mu)}{\pi}
\left(\frac{\pi^{2}}{3}-\frac{20}{3}\right)\right]\nonumber
\\[2mm]
\Gamma_{had}&=&\Gamma_{had}^{Born}
\left[1+\frac{\alpha_{s}(\mu)}{\pi}
\left\langle \left(11-\frac{2}{3}n_{f}\right)
\ln\frac{\mu}{M}
+\frac{159}{6}-\frac{31}{24}\pi^{2}
-\frac{8}{9}n_{f}
\right\rangle\right]
\label{cott}
\\[2mm]
\Gamma_{t\bar{t}}&=&\Gamma_{t\bar{t}}^{Born}
\left(1-4\frac{\alpha_{s}(\mu)}{\pi}\right)
\left(1+\frac{4}{3}\frac{\alpha_{s}(\mu)}{\pi}
\frac{1}{\beta_{t}}
\left[A(\beta_{t})+P(\beta_{t})\ln\frac{1+\beta_{t}}{1-\beta_{t}}
+Q(\beta_{t})
\right]\right)\nonumber
\end{eqnarray}
where $n_{f}=6$ denotes the number of light flavours (including top).
In the last equation
of (\ref{cott}) the functions $A(\beta_{t}), P(\beta_{t})$ and
$Q(\beta_{t})$  are given by
\begin{eqnarray}
A(\beta_{t})&=& (1+\bt^2)\left[
\frac{\pi^{2}}{6}+\ln\frac{1+\bt}{1-\bt}\,
\ln\frac{1+\bt}{2}
+2 \mbox{Li}_{2}\left(\frac{1-\bt}{1+\bt}\right)
+2 \mbox{Li}_{2}\left(\frac{1+\bt}{2}\right)\right.
       \nonumber\\
&-&\hspace{-2mm} \left.2 \mbox{Li}_{2}\left(\frac{1-\bt}{2}\right)
   -4 \mbox{Li}_{2}(\bt)
   +\mbox{Li}_{2}(\bt^{2})\right]
   +3\bt\ln\frac{1-\bt^{2}}{4\bt}-\bt\ln \bt .\\[2mm]
P(\beta_{t})&=&
\frac{19}{16}+\frac{2}{16}\bt^{2}+\frac{3}{16}\bt^{4}\\[2mm]
Q(\beta_{t})&=&\frac{21}{8}\bt-\frac{3}{8}\bt^{3}
\end{eqnarray}
where
$\beta_{t}=\sqrt{1-4m_{t}^{2}/M^{2}}$.
The first factor in the corrections  for $t\bar{t}$
has its origin in the $\eta_{Q}$ vertex correction, the second
factor is given in \cite{Reinders}.
Finally, the scale $\mu$ in $\alpha_{s}$ is adopted to M
 in our numerical calculation.
The QCD corrections to the $\gamma\gamma$ decay rate are negative
at the level of about -10\% whereas the corrections to the hadronic
decay increases the decay rate by about +35\%.
The QCD corrections to $\Gamma_{t\bar{t}}$ depend on the mass
of the boundstate and vary from +40\% for $M_{\eta}=300$ GeV
to -20\% for $M_{\eta}=2000$ GeV.

Since $\gamma\gamma, \gamma Z,ZZ$ and $WW$ can be identified in
the limit $M_{Z},M_{W}<<2m_{b',D}$
 after appropriate adjustment of the gauge
coupling, the QCD corrections can be assumed to cancel to a
large extent in the ratios listed in eqs. (\ref{dgamz}, \ref{dzz},
\ref{dww}).
Corrections for
the HZ mode are missing and we shall simply adopt eq. (\ref{dhz}) for
the present purpose.  The normalized decay rates $R_{ab}$  are
shown in figs. 4a and b for $\eta_{b^{\prime}}$ and $\eta_{D}$
respectively, the resulting branching ratios
$\Gamma_{ab}/\sum\Gamma_{ab}$
are displayed
in figs. 4c and d.  The pattern is significantly different for
the two cases:

The branching ratio of $\eta_{b^{\prime}}$ into $HZ$ is sizable and
 dominant for a large mass range.  Also the $WW$ mode is
fairly important, amounting to about 1\%  of the hadronic
mode, which is presumably difficult to detect
experimentally.
The $\gamma\gamma,ZZ$ and $\gamma Z$
modes are all relatively small
with their ratio basically governed by their relative
couplings and the statistical factor 1/2 for
$\gamma\gamma$ and $ZZ$.
\begin{equation}
\Gamma_{WW}\,:\,\Gamma_{ZZ}\,:\,\Gamma_{\gamma Z}\,:\,
                    \Gamma_{\gamma\gamma}
\approx
1 \,:\, 0.462 \,:\, 0.064 \,:\, 0.021
\end{equation}
This expectation for asymptotic quark mass
values is well met above
$m_{b^{\prime}}> 150 \,\mbox{GeV}$.
One thus anticipates a  large  production rate for $HZ$ and
even $WW$ will become fairly important.

This is in contrast to the situation for $\eta_{D}$ decays:  $HZ$ and
$WW$ are absent, and the asymptotic relation reads
\begin{equation}
\Gamma_{\gamma\gamma}\,:\,\Gamma_{\gamma Z}\,:\, \Gamma_{ZZ}
\approx
1 \,:\,  0.60 \,:\, 0.091
\end{equation}
In
this case $\gamma\gamma,\,\gamma Z$
and $ZZ$ constitute the dominant visible
modes with a
branching ratio of about $10^{-4}$.

  The resulting product of
production cross sections times branching ratio
for SSC and LHC energies
 are displayed
in figs. 5a and b for several interesting
$\eta_{b^{\prime}}$ channels, indicating
large rates in particular for $HZ$.
The production and decay of $\eta_{D}$ is shown in figs. 5c and d
indicating that $\eta_{D}$  could be
observed in the  $\gamma\gamma,\, \gamma Z$ and $ZZ$ modes
 with
characteristic signals.\\[5mm]
To summarize:\\
Production and decay rates for bound states of exotic
quarks can be predicted unambigously.
The uncertainty from the QCD potential, $\alpha_{s}$
and the structure functions appears to be well
under control. The rates for $pp\rightarrow \eta_{b^{\prime}}
(\rightarrow HZ)\, + \,X$ are particularly
promising  as a consequence of the large branching ratio.
The branching rations into $WW,\,ZZ,\,\gamma Z,$ or $\gamma\gamma$
are significantly smaller but in any case larger than the
corresponding branching ratios of a toponium resonance.

Throughout this paper the SQD mode has been ignored completely,
a clearly oversimplifying assumption.
The mixing of both $b'$ and $D$ with other quarks
is presumably small and it seems difficult to make
any hypothesis.
Instead we compare the total annihilation rate (for
$V_{J}$ with $\Lms{4}=300$ MeV) with the single quark
decay rate (through the charged current)
in fig. 6
 for an assumed mixing angle
of $10^{-2}$. This illustrates that annihilation decays would
be clearly dominant for $b'$ under this assumption.
For a complete treatment of $\eta_{D}$ we would have to include
FCNC single quark decays $D\rightarrow qZ$.
In view of the completely unknown mixing angle we refrain from a
detailled analysis. It is evident from fig. 6
that for mixing angles of $O(10^{-2})$ or less and $m_{D}$ below
$\sim 500$ GeV the main conclusions remain unaffected.

\def\app#1#2#3{{\it Act. Phys. Pol. }{\bf B #1} (#2) #3}
\def\apa#1#2#3{{\it Act. Phys. Austr.}{\bf#1} (#2) #3}
\def\lhc{Proc. LHC Workshop, CERN 90-10}
\def\npb#1#2#3{{\it Nucl. Phys. }{\bf B #1} (#2) #3}
\def\plb#1#2#3{{\it Phys. Lett. }{\bf B #1} (#2) #3}
\def\prd#1#2#3{{\it Phys. Rev. }{\bf D #1} (#2) #3}
\def\prl#1#2#3{{\it Phys. Rev. Lett. }{\bf #1} (#2) #3}
\def\prc#1#2#3{{\it Phys. Reports }{\bf C #1} (#2) #3}
\def\cpc#1#2#3{{\it Comp. Phys. Commun. }{\bf #1} (#2) #3}
\def\nim#1#2#3{{\it Nucl. Inst. Meth. }{\bf #1} (#2) #3}
\def\pr#1#2#3{{\it Phys. Reports }{\bf #1} (#2) #3}
\def\sovnp#1#2#3{{\it Sov. J. Nucl. Phys. }{\bf #1} (#2) #3}
\def\jl#1#2#3{{\it JETP Lett. }{\bf #1} (#2) #3}
\def\jet#1#2#3{{\it JETP Lett. }{\bf #1} (#2) #3}
\def\zpc#1#2#3{{\it Z. Phys. }{\bf C #1} (#2) #3}
\def\ptp#1#2#3{{\it Prog.~Theor.~Phys.~}{\bf #1} (#2) #3}
\def\nca#1#2#3{{\it Nouvo~Cim.~}{\bf #1A} (#2) #3}
\newpage
\sloppy
\raggedright

\vspace{1cm}
\noindent
{\bf Figure captions}\\[2mm]
\begin{itemize}
\item[{\bf Fig. 1}]
 Predictions for
$R(0)^2/M^3$ of the S--wave ground state
as a  function of $M$ for the potential $V_{J}$ with
$\Lms{4}$=200, 300 and 500 MeV (dotted, solid and dashed lines).
\item[{\bf Fig. 2}]
Ratio between the radiatively corrected production
 cross section
 for $pp\rightarrow \eta_{b^{\prime},D}+X$
and the lowest order result
for $\sqrt{S}$=16 and 40 TeV.
 We use MT set B1 parton distributions with
$\Lms{4}$=194 MeV  and work in the DIS factorization
scheme.
\item[{\bf Fig. 3}]
a) Production
 cross section
 for $pp\rightarrow \eta_{b^{\prime},D}+X$ in NLO
for $\sqrt{S}$=16 and 40 TeV.
 We use MT set B1 parton distributions with
$\Lms{4}$=194 MeV  and work in the DIS factorization
scheme (potentials as in fig. 1).\\
b) Dependence of the cross section on the parton distribution
   functions: MT set B1 \cite{MT} (solid),
   MT set SN \cite{MT} (dotted),  MRS set B200 \cite{MRS}
   (short-dash-dotted),
   KMRS set B \cite{KMRS} (long-dash-dotted)  and
   GRV \cite{GRV} (dashed).
\item[{\bf Fig. 4}]
Ratios $R_{ab}=\Gamma_{ab}/ \Gamma_{\gamma\gamma}$
 ($ab=HZ,t\bar{t},gg,WW,ZZ,\gamma Z$)
of $\eta_{b^{\prime}}$ (fig. a) and
 $\eta_{D}$ (fig. b)
 as functions of $M$
 and corresponding branchings ratios (figs. c and d)
for a top mass of 140 GeV in $\Gamma_{t\bar{t}}$.
\item[{\bf Fig 5.}]
 Cross section for $\eta_{b^{\prime}}$ (fig. a,b)  and
$\eta_{D}$ (fig. c,d)  production
including NLO corrections
multiplied by the  branching ratios
at $\sqrt{S}=40$ TeV and 16 TeV.
 We use MT set B1 parton distributions with
$\Lms{4}$=194 MeV  and work in the DIS factorization
scheme (potentials as in fig. 1).
\item[{\bf Fig 6.}]
Total widths for $\eta_{b', D}$ from annihilation (see fig. 4)
compared with the widths from single quark decays into
$b+W$.
\end{itemize}
\end{document}